\documentclass[aps, prl, preprint, superscriptaddress]{revtex4}
\usepackage{graphicx}
\usepackage{dcolumn}
\usepackage{bm}
\usepackage{natbib}

\begin{document}

\title{Negative refraction in natural ferromagnetic metals}

\author{S. Engelbrecht}
\author{A. M. Shuvaev}
\affiliation{Experimentelle Physik IV, Universit\"{a}t W\"{u}rzburg,
97074 W\"{u}rzburg, Germany} %
\author{Y. Luo}
\author{V. Moshnyaga}
\affiliation{I. Physikalisches Institut,
Universit\"{a}t G\"{o}ttingen, 37073 G\"{o}ttingen, Germany} %
\author{A. Pimenov}
\affiliation{Experimentelle Physik IV, Universit\"{a}t W\"{u}rzburg,
97074 W\"{u}rzburg, Germany} %

\date{\today}

\begin{abstract}
It is generally believed that Veselago's criterion for negative
refraction cannot be fulfilled in natural materials. However,
considering imaginary parts of the permittivity ($\varepsilon$) and
permeability ($\mu$) and for metals at not too high frequencies the
general condition for negative refraction becomes extremely simple:
$Re(\mu) < 0 \Rightarrow Re(n) < 0$. Here we demonstrate
experimentally that in such natural metals as pure Co and FeCo alloy
the negative values of the refractive index are achieved close to
the frequency of the ferromagnetic resonance. Large values of the
negative refraction can be obtained at room temperature and they can
easily be  tuned in moderate magnetic fields.
\end{abstract}

\maketitle
%\section{introduction}
The topic of negative refraction \cite{veselago.1968} has attracted
much research interest in the last years
\cite{Pendry.2004_1,Smith.2004}. Various possible realizations of a
negative index material has been proposed
\cite{Smith.2000,cubukcu_nature_2003,Pendry.2004_2,Pimenov.2005,Podolskiy.2005,Pimenov.2007,valentine_nature_2008}.
Experimental demonstrations of negative refraction in metamaterials
\cite{Shelby.2001,padilla.2006} and multilayers \cite{Pimenov.2005}
are based on the classical criterion by Veselago which requires
simultaneous negativity of electric permittivity $\varepsilon$ and
magnetic permeability $\mu$ \cite{veselago.1968}. However, natural
materials with these properties most probably do not exist. An
alternative way to achieve negative refraction in natural material
may try to use the extended criterion
\cite{McCall.2002,Boardman.2005} which takes into account imaginary
parts of the permittivity and permeability:
\begin{equation}
(\varepsilon_1 + \vert \varepsilon^{*} \vert)(\mu_1 + \vert \mu^{*}
\vert) < \varepsilon_2 \mu_2. \label{negcrit}
\end{equation}
This equation is straightforwardly derived from the inequality
$Re(\sqrt{\varepsilon^* \mu^*})< 0$. Here $\varepsilon^{*} =
\varepsilon_1 + i \varepsilon_2$ denotes the complex dielectric
permittivity and $\mu^{*} = \mu_1 + i \mu_2$ the complex magnetic
permeability, respectively. Eq. (\ref{negcrit}) has an interesting
consequence if we consider the electrodynamics of metals
\cite{dressel_book}. For metals at frequencies far below the
scattering rate the imaginary part of the permittivity dominates:
 $\varepsilon_1 << \varepsilon_2$ and, therefore,
$\vert \varepsilon{^*} \vert \approx \varepsilon_2$. In this case
Eq. (\ref{negcrit}) can be transformed to a simple final condition
for the negative refraction: $\mu_1 < 0 \Rightarrow Re(n^*)<0$ .
This condition should be fulfilled in ferromagnetic metals close to
the ferromagnetic resonance if the strength of the mode is high
enough. First experimental demonstration of negative refraction in a
ferromagnetic metal \cite{Pimenov.2007} utilized the ferromagnetic
resonance in a Ca doped LaMnO$_3$. Although a sufficiency of the
simple condition $\mu_1 < 0$ for metals has been proven, the
existence of natural materials with negative refraction still had to
be demonstrated.

In this Letter we present the results of the refractive index
experiments in real natural metals. As examples of natural
ferromagnetic metals we have chosen pure Cobalt and Fe/Co alloy.
Using polarization and phase controlled experiments in the
millimeter wave range we demonstrate that close to the frequency of
the ferromagnetic resonance the refractive index of Co and FeCo
alloy indeed goes deep into the negative regime.
%\section{experimental}

Pure Cobalt and Fe$_{0.5}$Co$_{0.5}$ alloy were prepared as
polycrystalline thin films by magnetron sputtering technique. The
thickness of both samples was $150\pm 30$\,nm. As substrate MgO was
used, whose millimeter wave properties have been determined in a
separate experiment as n$_{MgO} = (3.09\pm
0.01)+1.4\cdot10^{-9}(T[K])^3$. Here the temperature $T$ is given in
Kelvin. The imaginary part of the refractive index of the MgO
substrate in the frequency range of the present experiment was below
$1\cdot10^{-4}$ at all temperatures.

\begin{figure}
\centering
\includegraphics[width=0.7\linewidth, clip]{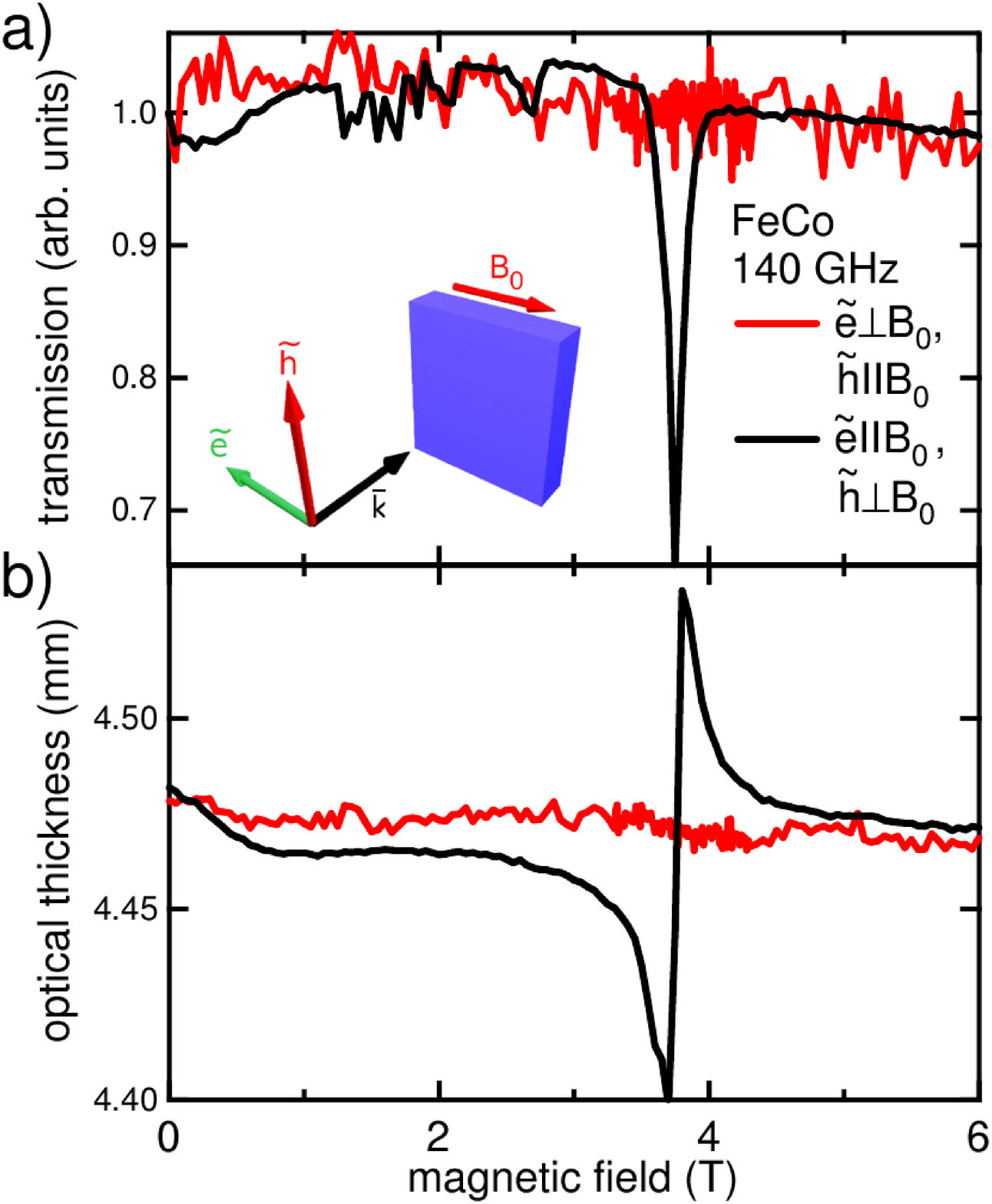}
\caption{ a) transmittance and b) phase shift of FeCo alloy for two
different excitation geometries at T=170$\,$K. Within the geometry
$\tilde{h} \perp B_0$ the ferromagnetic resonance can be excited,
whereas for $\tilde{h} \vert \vert B_0$ no excitation occurs. The
inset sketches the magnetically active geometry ($\tilde{h} \perp
B_0$). Here $\bar{k}$ denotes the wave vector, $\tilde{h}$ and
$\tilde{e}$ are the ac magnetic and electric fields of the
radiation, and B$_0$ is the static external magnetic field.}
\label{excitationgeo}
\end{figure}

The experiments in the millimeter frequency range were carried out
in a Mach-Zehnder interferometer setup \cite{Pimenov.2005_2,
Kozlov_book} using backward wave oscillators as radiation source.
This spectrometer enables to measure both the transmittance and
phase shift as function of frequency, temperature or external
applied magnetic field within controlled polarization geometries.
Magnetic field experiments were carried out using a split coil
magnet with polypropylene windows.

The experimental spectra obtained were analyzed using the Fresnel
formulas for the complex transmission coefficient of the
substrate-film system \cite{Pimenov.2005_2, Kozlov_book,Born.2006}.
A non-trivial problem of the experiment is the separation of
dielectric and magnetic properties of the sample. In this case four
independent experimental values are necessary. In present
experiments, we obtained the transmittance and phase shift of the
sample within two different polarizations of the incident radiation,
$\tilde{h} \vert \vert B_0$ and $\tilde{h} \perp B_0$. Here
$\tilde{h}$ is the ac magnetic field of the radiation and $B_0$ is
the external magnetic field. From the solution of the Bloch's
equations for the magnetic moments in external magnetic fields it
follows that nonzero magnetic susceptibility can only be obtained in
the geometry $\tilde{h} \perp B_0$ \cite{Abragam.1970}. The inset of
Fig. \ref{excitationgeo} shows the geometry in which the
ferromagnetic resonance is excited. To determine the parameters of
the ferromagnetic resonance a Lorentz line shape was used, i.e. the
magnetic permeability was taken as
\begin{equation}
\mu^*(B) = 1 + \frac{\Delta \mu B B_0}{B^2 - B_0^2 - i B_0 \Gamma}.
\label{lorentz}
\end{equation}
Here $\Gamma$ is the width of the resonance, $B_0$ the resonance
field and B the external applied magnetic field.

\begin{figure}
\centering
\includegraphics[width=0.9\linewidth, clip]{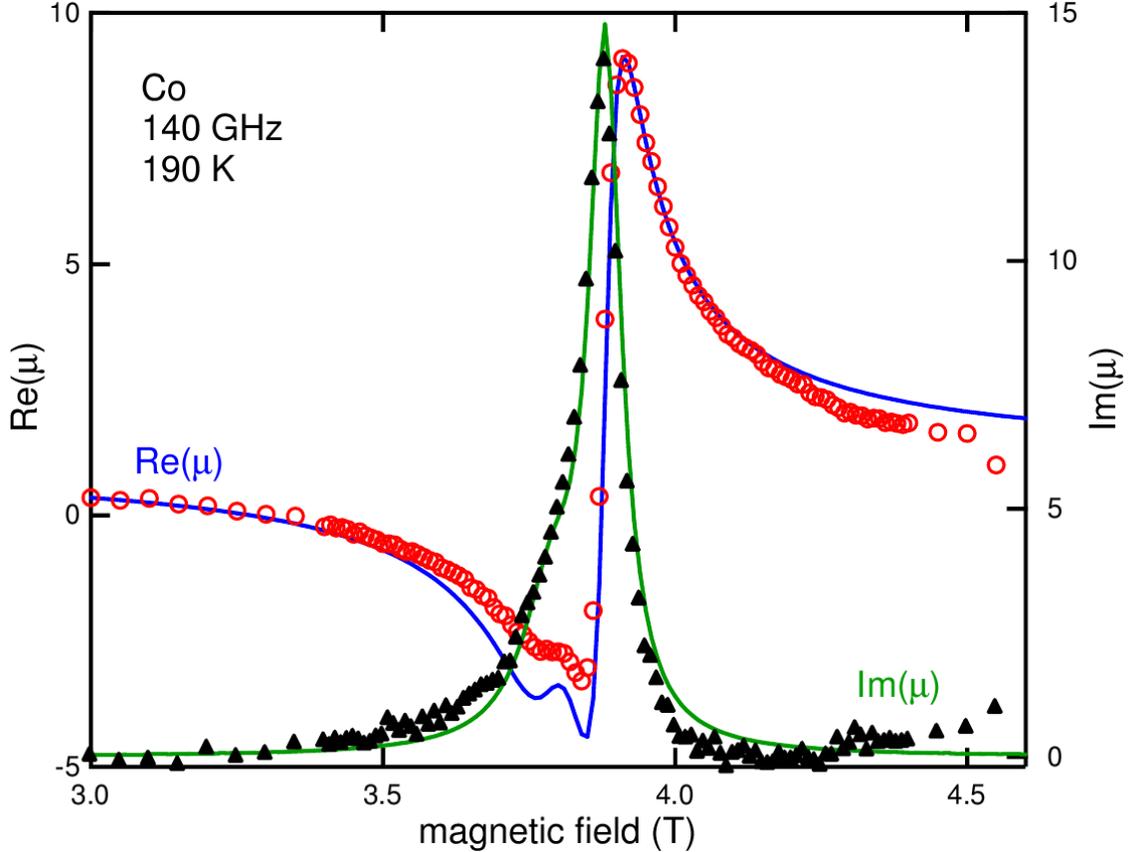}
\caption{ Complex magnetic permeability obtained from transmission
and phase shift data (symbols) and fitted using Eq. \ref{lorentz}
(solid lines). Close to the resonance the real part of the
permeability becomes negative.} \label{muco}
\end{figure}

Fig. \ref{excitationgeo} shows the transmittance and phase shift of
the FeCo alloy at T=170$\,$K and at a frequency of 140$\,$GHz in
both excitation geometries. As can be clearly seen in the geometry
$\tilde{h} \vert \vert B_0$ no excitation of the ferromagnetic
resonance occurs ($\mu^*$ =1) and, therefore, we can use this
geometry to determine the dielectric properties of the metal films.
These properties are obtained from the absolute transmittance
spectra in zero field and are well described by conventional Drude
conductivity of a metal in the millimeter range (Table
\ref{fitparam}).

Furthermore, in the nonmagnetic geometry  no magnetoresistance
occurs within our experimental resolution neither in the FeCo alloy
nor in pure Co. Therefore we can assume a magnetic field independent
behavior of the permittivity. In contrast, in the geometry
$\tilde{h} \perp B$ the resonance can be excited which allows to
independently determine the permeability. The excitation frequency
is dependent on the resonance field and follows the relationship
$\omega_{res} = \gamma \sqrt{(H_0 + 4 \pi M_0)H_0}$ due to the
geometry of the experiment (Voigt geometry) \cite{Zvezdin.1997}.
Here $\omega_{res}$ denotes the excitation frequency, $\gamma$ is
the gyromagnetic ratio and $M_0$ is the static magnetization.  Using
the dielectric permittivity obtained from the magnetically inactive
geometry we are able to calculate the magnetic permeability which is
shown in Fig. \ref{muco} for the Co sample at T=190$\,$K.

\begin{table}
\begin{center}
\caption{ Electrodynamic parameters for FeCo and Co samples at two
different excitation frequencies and temperatures.} \label{fitparam}
\begin{tabular}{c c c c c }
\hline \hline
sample          & \multicolumn{2}{c}{FeCo}          &    \multicolumn{2}{c}{Co}       \\
frequency[GHz]  &   140           &     182         &       140       & 180           \\
\hline
T = 10$\,$K     &               &   &   &   \\
B$_0$ [T]       & 3.73$\pm$0.01   & 5.12$\pm$0.01   & 3.87$\pm$0.01   & 5.20$\pm$0.01 \\
$\Delta \mu$    &   0.32$\pm$0.05   &   0.16$\pm$0.05   &   0.20$\pm$0.05   &   0.13$\pm$0.05 \\
$\Gamma$ [T]    & 0.07$\pm$0.02   & 0.08$\pm$0.02    & 0.08$\pm$0.02   & 0.11$\pm$0.02  \\
$\sigma_{}$[$10^6\,\Omega^{-1}m^{-1}$] & \multicolumn{2}{c}{8$\pm$2} & \multicolumn{2}{c}{12$\pm$2} \\
\hline
T = 300$\,$K    &   &   &   &   \\
B$_0$ [T]      & 3.76$\pm$0.01   & 5.17$\pm$0.01   & 3.89$\pm$0.01   & 5.21$\pm$0.01 \\
$\Delta \mu$    &   0.29$\pm$0.05   &   0.27$\pm$0.05   &   0.38$\pm$0.05   &   0.21$\pm$0.05 \\
$\Gamma$ [T]    & 0.08$\pm$0.02   & 0.1$\pm$0.02    & 0.07$\pm$0.02   & 0.11$\pm$0.02  \\
$\sigma_{}$[$10^6\,\Omega^{-1}m^{-1}$] & \multicolumn{2}{c}{6$\pm$1} & \multicolumn{2}{c}{6$\pm$1} \\
\hline \hline
\end{tabular}
\end{center}
\end{table}

%\section{discussion}
Table \ref{fitparam} summarizes the film properties for FeCo and Co
at two different excitation frequencies. For both samples we get
quite similar electrodynamic parameters. Weak temperature dependence
of the resonance fields is due to the decreasing static
magnetization with increasing temperature. These effects are quite
small, since the Curie temperatures of both Co and FeCo are far
above room temperature (Co: T$_C$=1390$\,$K \cite{Mohn.1987}, FeCo:
T$_C$=1253$\,$K \cite{James.1992}).

\begin{figure}
\centering
\includegraphics[width=0.7\linewidth, clip]{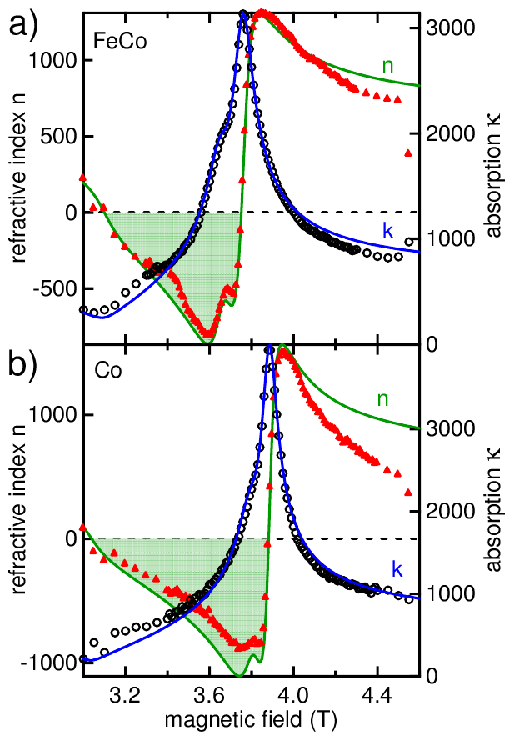}
\caption{ Magnetic field dependent refractive index for the a) FeCo
alloy and b) metallic Co at T=300$\,$K and at 140$\,$GHz. The shaded
area denotes the range in which the real part of the refractive
index becomes negative. Symbols - experimental data, solid lines are
fits using Eq. (\ref{lorentz})} \label{bla}
\end{figure}

With the dielectric permittivity and the magnetic permeability
determined, we can now calculate the refractive index by n =
$\sqrt{\varepsilon^* \mu^*}$. Here we note that due to a continuous
variation of the refractive index as function of external magnetic
fields the sign of the square root is obtained automatically on
crossing the imaginary axis. Fig. \ref{bla} a) and b) show the
magnetic field dependence of the complex refractive index of the
FeCo and Co sample, respectively. The shaded area denotes the range
in which the refractive index becomes negative. Here we see that
even at room temperature these metals show a negative index of
refraction, which is also highly tunable by the applied magnetic
field. We note that the second weak resonance which can be seen in
these plots (at $\sim$3.7$\,$T for FeCo and $\sim$3.8$\,$T for Co)
most probably arises due to inhomogeneities in the film
magnetization \cite{Semeno.2009}. The estimate of the best value of
the figure of merit for our samples gives $|n|/\kappa \sim 0.8$.
This value is already close to the theoretical limit $|n|/\kappa =
1$ for ferromagnetic metals.

A remaining problem with the present materials is the high
absorption coefficient in the region of negative refraction.  High
absorption values leads to small transmissions of the order of
$10^{-4}$ and have to be overcome for possible applications. A
possible solution to this problem could be the usage of either
superconductors or very clean metals. In these cases and in the
millimeter range the conductivity will be purely imaginary, i.e. no
internal absorption will be present. Unfortunately, at current stage
of research both solution would be difficult, as e.g. magnetism and
superconductivity tend to exclude each other.

%\section{conclusion}
In conclusion, we have shown that the refractive index in natural
ferromagnetic metals becomes negative close to the frequency of the
ferromagnetic resonance. We have investigated pure Co and FeCo alloy
both revealing negative values of the refractive index over a large
temperature range and at different excitation frequencies. Both
metals show large negative index of refraction at room temperature
which is highly tunable by the external magnetic field.

\bibliography{metamaterials}{}

\end{document}